\documentclass[sigconf]{acmart}

\usepackage{lineno}
\usepackage{algorithmic}
\usepackage{textcomp}
\usepackage{xcolor}
\graphicspath{{images/}}
\usepackage{subfigure}
\usepackage{tikz} %
\usepackage[capitalize,noabbrev]{cleveref} %
\usetikzlibrary{decorations.pathreplacing}
\usepackage[group-separator={,},group-minimum-digits=4]{siunitx}
\usepackage[inline]{enumitem}
\usepackage[multiple]{footmisc} %
\usepackage{csquotes} %
\usepackage{float}
\usepackage{balance}

\newenvironment{boxed_new}
    {\begin{center}
    \begin{tabular}{|p{0.45\textwidth}|}
    \hline\\
    }
    { 
    \\\\\hline
    \end{tabular} 
    \end{center}
    }

\newcommand{\cor}{code review}
\newcommand{\gh}{{G}it{H}ub}
\newcommand{\fbsd}{{F}ree{BSD}}
\newcommand{\llvm}{\textsc{LLVM}}
\newcommand{\mcr}{{M}odern {C}ode {R}eview}
\newcommand{\mo}{{M}ozilla}
\newcommand{\oss}{\textsc{OSS}}
\newcommand{\tta}{time-to-accept}
\newcommand{\ttfr}{time-to-first-response}
\newcommand{\ttm}{time-to-merge}
\newcommand{\ger}{{G}errit}
\newcommand{\phab}{{P}habricator}
\newcommand{\npt}{non-productive time}

\newcommand{\initialcrcount}{\num{569914}}
\newcommand{\finalcrcount}{\num{350043}}
\newcommand{\initialgercrcount}{\num{329392}}
\newcommand{\initialphabcrcount}{\num{240522}}
\newcommand{\filteredgerreviews}{\num{164726}}
\newcommand{\filteredphabreviews}{\num{185317}}

\hypersetup{draft}

\acmConference[MSR 2022]{MSR '22: Proceedings of the 19th International Conference on Mining Software Repositories}{May 23–24, 2022}{Pittsburgh, PA, USA}

\begin{document}

\title{Mining Code Review Data to Understand Waiting Times Between Acceptance and Merging: An Empirical Analysis}

\begin{abstract}
Increasing code velocity (or the speed with which code changes are reviewed and merged) is integral to speeding up development and contributes to the work satisfaction of engineers.
While factors affecting code change acceptance have been investigated in the past, solutions to decrease the code review lifetime are less understood.
This study investigates the code review process to quantify delays and investigate opportunities to potentially increase code velocity.
We study the temporal characteristics of half a million code reviews hosted on Gerrit and Phabricator, starting from the first response, to a decision to accept or reject the changes, and until the changes are merged into a target branch. 
We identified two types of time delays:
(a) the wait time from the proposal of code changes until first response, and
(b) the wait time between acceptance and merging.
Our study indicates that reducing the time between acceptance and merging has the potential to speed up Phabricator code reviews by 29--63\%. 
Small code changes and changes made by authors with a large number of previously accepted code reviews have a higher chance of being immediately accepted, without code review iterations. 
Our analysis suggests that switching from manual to automatic merges can help increase code velocity. 
\end{abstract}

\author[1]{Gunnar Kudrjavets}
\affiliation{%
   \institution{University of Groningen}
   \city{Groningen}
   \country{Netherlands}}
\email{g.kudrjavets@rug.nl}

\author[2]{Aditya Kumar}
\affiliation{
    \institution{Snap, Inc.}
    \streetaddress{2772 Donald Douglas Loop N}
    \city{Santa Monica}
    \state{CA}
    \country{USA}
    \postcode{90405}}
\email{adityak@snap.com}

\author[3]{Nachiappan Nagappan}
\affiliation{
    \institution{Facebook, Inc.}
    \streetaddress{1 Hacker Way}
    \city{Menlo Park}
    \state{CA}
    \country{USA}
    \postcode{94025}}
\email{nnachi@fb.com}

\author[4]{Ayushi Rastogi}
\affiliation{%
   \institution{University of Groningen}
   \city{Groningen}
   \country{Netherlands}}
\email{a.rastogi@rug.nl}



\maketitle

\section{Introduction}

\emph{Code velocity} or the speed with which code changes are reviewed and merged is an important industry metric~\cite{savor_2016,kononenko_2016}.
Increased code velocity also contributes to an engineer's job satisfaction~\cite{savor_2016,kononenko_2016}.
In studies and surveys, engineers repeatedly point to delays in response times as pain
points in their evaluation of the efficiency of \cor\ process~\cite{bird_2015,macleod_2018}.

Existing studies that analyze the \cor\ process focus primarily on improving
one aspect of the \cor\ life cycle: the time it takes for the proposed changes
to be accepted by a reviewer~\cite{beller_2014,rigby_2015}.
For example, there are studies that examine how code size impacts the acceptance time~\cite{jiang_will_2013,kononenko_2016,weisgerber_small_2008,tsay_influence_2014}. 
These studies systematically ignore potential inefficiencies that exist elsewhere in the \cor\ process which, if removed, can increase code velocity perhaps at even lesser cost. 

Our experience working at X and Y (names omitted in the peer review version to comply
with the guidelines for double-blind review) suggests that often the proposed
code changes sit idle with no response or action.
This time spent idling with no productive action (also referred to herein as \emph{\npt}) can happen during the different phases of \cor.
Understanding where and why such \npt s occur can point to solutions to increase code velocity.
Particularly, we investigate:

\begin{quote}
    \textbf{\textsc{RQ1}}: Does \npt\ exist in the \cor\ process? If so, how much?
\end{quote}

We define the distinct phases of \cor\ process starting from proposing a change, the first response, decision to accept or reject a change, until integration into target branch. 
We report descriptive statistics to quantify the extent of the problem. 
Contingent on the presence of  \npt, we explore:

\begin{quote}
	\textbf{\textsc{RQ2}}: What contributes to \npt\ in the \cor\ process? Can we reduce it? 
\end{quote}
Building on the insights from \textsc{RQ1} and our industry experience, we
investigate what happens to the \cor s after they are accepted and explore a way to minimize \cor\ iterations.

We collected data about \initialcrcount \cor s from \ger\ and \phab~\cite{gerrit_proper,phabricator_proper} to quantify the phases of \cor\ process and activities within and examine the \npt\ for potential solutions. 
The choice of the platforms is motivated by (a) the popularity of projects hosted on these platforms (e.g., \mo\ and Eclipse), (b) availability of archival data for investigation that is otherwise not possible on platforms like \gh, and (c) the distinctive processes of the two platforms for comparison.   
To the best of our knowledge, this is the first study that quantifies non-productive time in Gerrit and Phabricator code reviews on a large scale.

Our study identifies two areas of time delays in the code review process: 
(a) from the time a change is proposed until it gets the first response, and 
(b) time between acceptance and integration.
Our study suggests that the smaller size of code change and author's prior successful code review experience speeds up the first response time, potentially influencing the overall integration time. 
Further, our findings indicate that switching to an automatic merge-on-acceptance model can potentially reduce the overall \ttm\ by eliminating \npt.

\section{Background and Related Work}

\subsection{Motivation}
\mcr\ sets high expectations for \cor\ times and requires
them to be completed in a timely manner.
Existing research indicates that \cor s on Google, Microsoft, and open-source software (\oss) projects take approximately 24 hours~\cite{rigby_2013}.
Google's recommended \cor\ best practices state explicitly that
\textquote{we expect feedback from a code review within 24
(working) hours}~\cite[p.~176]{winters_2020}.
Chromium \cor\ guidelines recommend that \textquote{aim to provide some
kind of actionable response within 24 hours of receipt (not counting weekends and holidays)}~\cite{chromium_2021}.
From the projects we investigate in this paper, \fbsd\ and \llvm\ do not
specify fixed deadlines~\cite{freebsd_cr_guidance_2021, llvm_cr_guidance_2021}.
However, \mo\ recommends that \textquote{strive to answer in a couple of days, or at least
under a week}~\cite{mozilla_cr_guidance_2021}.
Blender sets the expectation that \textquote{[d]evelopers are expected to
reply to patches [in] 3 working days}~\cite{blender_cr_guidance_2021}.
Based on our industry experience at X and Y (names omitted in the peer review version to comply
with the guidelines for double-blind review), getting
the first response on the same day is highly desirable, with the \emph{24-hour
\cor\ completion time being a default expectation}.

\emph{Code velocity}, or the speed with which code changes are reviewed
and merged into the target branch, is an important metric
both in the industry and \oss\ community.
Facebook's experience shows that increased code velocity leads to
earlier detection of code defects and faster product deployments~\cite{savor_2016}.
Survey of \mo\ developers finds that quick \emph{turnaround time}
is important for both reviewers and engineers submitting the code
changes~\cite{kononenko_2016}.
In some cases, being less thorough during \mo\ \cor s is an acceptable
trade-off if it results in faster \cor s.
CodeFlow Analytics is a system at Microsoft that collects \cor\ data
and generates metrics from that data~\cite{codeflow_discussion}.
Researchers analyzing the usage of CodeFlow Analytics find that every
individual they talked to was interested in two code velocity related
points in time:
(a) when the \emph{first comment} or \emph{sign-off} from a reviewer occurs, and
(b) when the \cor\ has been marked as \emph{completed}~\cite{bird_2015}.
A study investigating the declining performance of \cor s in Xen hypervisor project finds that \emph{time-to-merge} is
the most important metric to understand the delays imposed
by the \cor\ process~\cite{izquierdo-cortazar_2017}.
Another study from Microsoft shows that when it comes to the challenges
faced during the code review, \emph{response time} is the number one
concern because \textquote{authors of code changes discussed how it's hard getting
feedback in a timely manner}~\cite{macleod_2018}.

Anecdotally, we observe that not responding to \cor s in a
timely manner causes frustration amongst engineers
attempting to contribute to the project.\footnote{\protect\url{https://reviews.llvm.org/D9766}}

\begin{quotation}
``I do think I have proven to be sticking around and can be trusted to continue work on that in the future. inf act, the process has been so slow that pretty much anyone would have given up by now.

It's been a month that this is just sitting there. You guys need to provide a way forward to get it as it is. I just can't reimplement that in a new way every 2 month because nobody has any clue how to move forward on the subject. This is disrespectful of my work and hurtful for the project at large as you can't get a better strategy to get contributor pissed of.''
\end{quotation}

Delaying the feedback also hinders the ability of a project to recruit new collaborators.\footnote{\protect\url{https://reviews.llvm.org/D40988}}

\begin{quotation}
``Nobody IS able to make changes, but not because of complexity!

We discourage away potential contributors/maintainers by leaving their reviews for weeks/months/years, not just not letting them in, not even discussing them..''
\end{quotation}

\subsection{Related work}

\subsubsection{Literature reviews}

There has been little focus on the efficiency of code reviews in comparison to the wide range of studies on code review process.
A systematic literature review on  \mcr\ (in the year 2021)
identifies only 4 papers related to \cor\ time out of the 139 studies
used for detailed analysis~\cite{davila_2021}.
The literature classifies review time into three categories:
\begin{enumerate*}[label=(\roman*),before=\unskip{ }, itemjoin={{, }}, itemjoin*={{, and }}]
    \item \emph{review delay} (time from the first review request submission to the first reviewer feedback)
    \item \emph{review duration} (time from the first review request submission to the review conclusion)
    \item \emph{review speed} (rate of \textsc{SLOC} reviewed per hour).
\end{enumerate*}
Another systematic literature review
of \mcr\ practices (in the year 2020) finds
only 3 papers (out of 51) on accelerating the software development
process~\cite{nazir_2020}.
A systematic mapping study (in the year 2019) shows that out of 177 research papers
categorized, only 9 focus on topics related to effectiveness and/or
efficiency of \mcr\ or integration delays of pull
requests~\cite{badampudi_2019}.

\subsubsection{Existing studies}

There are many definitions of \mcr.
One study defines \cor\ as \textquote{a process that takes as
input original source code (i.e., the first unreviewed change attempt),
and outputs accepted source code}~\cite{beller_2014}.
Another paper describes review time as \textquote{until the end of the discussion of the change}~\cite{rigby_2015}.

Most of the research relating to the effectiveness of \mcr\ has focused on how to
reduce the time it takes for changes to be reviewed, i.e., \tta.
Time-to-accept refers to the time when
changes are submitted for review until some sort of a conclusion is reached.
The prevailing wisdom is that engineers should attempt to submit
smaller changes because they are easier to understand and will solicit
actionable feedback~\cite{bacchelli_2013}.
Study on \mo\ project finds that developers \emph{feel}
that the size-related factors are the most important when 
it comes to code review time and decision~\cite{kononenko_2016}.
A study on the patch acceptance in Linux kernel found a
link between patch size and the time it takes to review the changes, \emph{but not
the overall patch integration time}~\cite{jiang_will_2013}.

We found very few studies dissecting the \cor\ lifetime into separate
intervals.
A study on \ger\ \cor s categorizes a \cor\ into
\emph{pre-review time}, \emph{review time}, \emph{pre-integration time},
and \emph{integration time}~\cite{lehtonen_2015}.
Pre-review time starts when change is uploaded to \ger\ and ends
when various builds and automated tests pass, and the author decides that
the changes are either suitable for reviewing or more work is needed
before another upload.
Review time starts immediately after that and ends when a decision
about the code changes has been reached.
The pre-integration time is defined as the time after the review is completed,
but the actual integration process is yet to start.
The integration time is a measure of how long it takes for the change
to be merged into the target branch after the review process is completed.

This implies that the code changes are not \textquote{real} until they are available for building, profiling, testing, and being executed in the production or test environments.
That is possible only after they have been merged into the main branch used for daily development.
Based on the above, the time until the code changes are integrated into a target branch seems a better metric.

\subsubsection{Terminology}
\label{subsubs:terminology}

\begin{figure*}[!htbp]
\centering
\begin{tikzpicture}[scale=0.8]
\filldraw 
(0,1) circle (2pt) node[align=left,   above] {Published} --
(0.95,1) circle (2pt) node[align=center, above] {}       --
(6,1) circle (2pt) node[align=center, above] {Accepted}  --
(9.1,1) circle (2pt) node[align=center, above] {Start of\\integration}     -- 
(15,1) circle (2pt) node[align=right,  above] {Merged};
\node[align=center] at (0.95,2.0) {Time-to-first-response};
\draw [-latex](0.95,1.8) -- (0.95,1.1);
\draw [thick,decorate,decoration={brace,amplitude=6pt,raise=0pt,mirror}] (0,0.75) -- (0.95,0.75);
\node[align=center] at (0.5,0.25) {\textbf{Non-productive time}};
\draw [thick,decorate,decoration={brace,amplitude=6pt,raise=0pt,mirror}] (0,-0.25) -- (1.9,-0.25);
\node[align=center] at (1,-0.75) {Revision $1$};
\draw [thick,decorate,decoration={brace,amplitude=6pt,raise=0pt,mirror}] (2,-0.25) -- (3.9,-0.25);
\node[align=center] at (2.95,-0.75) {\ldots};
\draw [thick,decorate,decoration={brace,amplitude=6pt,raise=0pt,mirror}] (4,-0.25) -- (6,-0.25);
\node[align=center] at (5,-0.75) {Revision $n$};
\draw [thick,decorate,decoration={brace,amplitude=6pt,raise=0pt,mirror}] (0,-1.25) -- (6,-1.25);
\node[align=center] at (3,-1.75) {Time-to-accept};
\draw [thick,decorate,decoration={brace,amplitude=6pt,raise=0pt,mirror}] (6.1,-1.25) -- (9,-1.25);
\node[align=center] at (7.5,-1.75) {\textbf{Non-productive time}};
\draw [thick,decorate,decoration={brace,amplitude=6pt,raise=0pt,mirror}] (9.1,-1.25) -- (15,-1.25);
\node[align=center] at (12,-1.75) {Integration time};
\draw [thick,decorate,decoration={brace,amplitude=6pt,raise=0pt,mirror}] (9.1,-0.25) -- (10.9,-0.25);
\node[align=center] at (10,-0.75) {Attempt $1$};
\draw [thick,decorate,decoration={brace,amplitude=6pt,raise=0pt,mirror}] (11,-0.25) -- (12.9,-0.25);
\node[align=center] at (11.95,-0.75) {\ldots};
\draw [thick,decorate,decoration={brace,amplitude=6pt,raise=0pt,mirror}] (13,-0.25) -- (15,-0.25);
\node[align=center] at (14,-0.75) {Attempt $m$};
\draw [thick,decorate,decoration={brace,amplitude=6pt,raise=0pt,mirror}] (0,-2.25) -- (15,-2.25);
\node[align=center] at (7.5,-3) {Time-to-merge};
\end{tikzpicture}
\caption{Code review life cycle (using \textsc{CI}) when the proposed changes are accepted for inclusion. A \cor\ can be abandoned by an author or rejected by a reviewer at any point during its life cycle.}
\label{fig:accepted_cr_timeline}
\end{figure*}
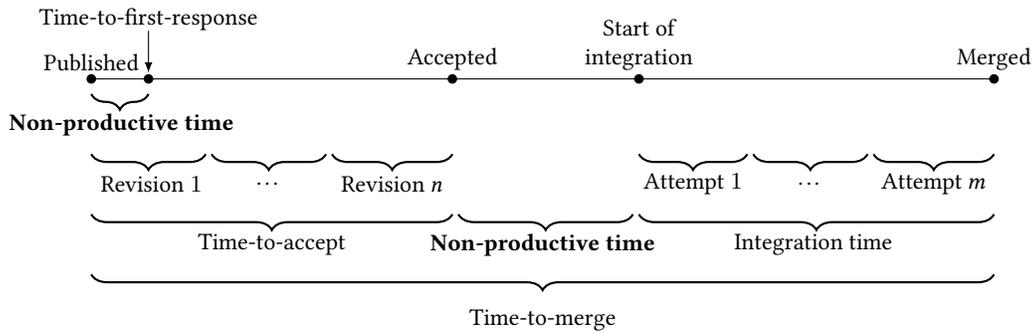

There are many ways to quantify code review time.
The term \emph{review interval} is defined as \textquote{the calendar
time since the commit was posted until the end of discussion of
the change}~\cite{rigby_2015}.
Another term utilized is \emph{time to completion} that is
defined as \textquote{the time between when the author submits
the review to when it has been marked as completed (usually indicating
that the change can be checked in)}.
Two studies use the term \emph{review window} meaning
\textquote{the length of time between the creation of
a review request and its final approval for
integration}~\cite{morales_2015,mcintosh_2016}.
The granularity of measurements is generally not specified, except
one paper which defines \emph{review length} as \textquote{time
in days from the first patch submission} and \emph{response delay}
as \textquote{time in days from the first patch submission to the
posting of the first reviewer message}~\cite{thongtanunam_2015}.
In addition, there is also a concept of \emph{resolve time} which
is defined as \textquote{the time spent from submission to the
final issue or pull request status operation
(stopped at \emph{committed}, \emph{resolved}, \emph{merged},
\emph{closed}) of a contribution}~\cite{zhu_2016}.

In the context of \gh\ and pull requests, the prevailing terms are
\emph{pull request latency} and \emph{pull request lifetime}, which
are respectively defined as \textquote{the time interval between pull request
creation and closing date} and \textquote{the time interval
between its opening and its closing}~\cite{yu_wait_2015,soares_2020}.
When investigating code velocity for \gh\ projects,
the term \emph{merge delay}
(days between when a pull request was submitted and when it was merged) is also used~\cite{bernardo_2018}.
\gh\ Insights adds two more additional terms to the mix: \emph{code
review turnaround} as \textquote{the time elapsed between a review assignment and a completed review} and \emph{time to merge} as \textquote{time between the first commit on a branch and the merge action of a pull request on that branch}~\cite{github_metrics_2021}.

\subsubsection{Metrics}
\label{subsubsec:metrics}
Inspired from the range of definitions available to describe code review time, we choose the following metrics to characterize \cor\ process: 

\begin{itemize}
    \item \textbf{Time-to-first-response}~\cite{bird_2015,macleod_2018}: is defined as the
    time from the publication of a code change until the first acceptance, comment,
    inline comment (comment on specific code fragments), or rejection by a person
    (excludes bots) other than the author of the code.
    
    \item \textbf{Time-to-accept}~\cite{bird_2015}: refers to the time from when a \cor\ is published
    for review until the acceptance of the \cor\ by someone other than the author of the code (excludes bots).
    
    \item \textbf{Time-to-merge}~\cite{github_metrics_2021,kononenko_2016}: is defined as
    \textquote{\ldots the time since the proposal of a change (\ldots) to the merging in the
    code base \ldots}~\cite{izquierdo-cortazar_2017}.
\end{itemize}

\cref{fig:accepted_cr_timeline} is a visual presentation of the relationship among these metrics.

\subsection{\mcr\ Life cycle}

Before Continuous Integration (\textsc{CI}) and Continuous Delivery (\textsc{CD})
became widely used approaches to releasing
software, the workflow for an engineer to commit accepted code changes
was relatively less time-consuming.
Upon receiving the acceptance to commit code changes,
an engineer must synchronize their local source tree with the latest changes,
build the merged changes,
re-run the relevant test cases,
commit the changes,
and push them to the target branch.
With \textsc{CI}/\textsc{CD} being the methodology of choice
for companies such as Amazon, Facebook, and Google, the amount of
validation done after code changes have been accepted has increased.
Introduction of \textsc{CI} tools such
as GitLab, Jenkins CI, Travis CI, added a
varying amount of validation steps to the process of merging the changes~\cite{duvall_continuous_2007}.
Each of those steps can increase the \ttm.

Based on our industry experience, the overall \cor\ timeline and its
composition is depicted in \cref{fig:accepted_cr_timeline}.
The \ttfr\ is a subset of the time it takes to complete the first iteration of the \cor.
The \tta\ contains all the iterations of the \cor\ until it is accepted.
The \ttm\ consists of \tta\ and the time it takes for the changes to be merged.

A major difference between industry and \oss\ projects is the presence
and scope of integration timeline.
Our experience suggests that the amount of verification done in industry is higher.
A modern industrial \textsc{CI} pipeline may consist of compiling the code changes
for multiple \textsc{CPU} architectures, operating systems, and use different compilers.
Validation steps such as linting code changes to verify conformance to coding standards,
using static analyzers to preemptively detect coding errors,
executing unit tests,
executing a set of functional tests specific to the area,
conducting performance experiments to avoid performance regressions,
and even performing a deployment of changes in an internal testing environment
are used~\cite{micco_2017, micco_2018,rossi_2017}.
Each of these steps can result in failures or abnormal results (indicating a
problem with the new code or validation suites themselves) that need to be
investigated and explained.
This can delay the merging of code changes.
Engineers need to either to manually investigate the failures or rely on automated retry mechanisms to repeat the validation.

\subsection{Overview of \ger\ and \phab}
\ger\ and \phab\ are code collaboration tools which operate using a similar workflow.
Engineers make code changes locally and when they decide that changes are
ready to be reviewed then they submit the changes using the respective
command-line or \textsc{GUI} tools.
Reviewers are either picked by an author, assigned automatically, or self-selected once the \cor\ becomes available.
Reviewers can perform a variety of actions on the \cor: accept, comment, reject, request new changes, etc.
Comments can target specific lines of code (in terms of \phab\ these are
called \emph{inline comments}) or just be general comments.
In \ger, an engineer can vote on a change by giving a response which by default
ranges from \textquote{-2 Do not submit} to \textquote{+2 Looks good to me, approved}.
By default, at least one \textquote{+2} vote without any \textquote{-2} votes is needed
before the changes can be merged.
\ger\ additionally introduces a separate label to signify that a change has been
\emph{verified} to work.
Verification is typically performed by an automated bot and consists of tasks
like compiling and linting the code, running unit tests or doing any other
basic verification steps specific to a project.
In \phab, the acceptance model is simpler and consists of actions such as acceptance,
rejection or requesting changes.
At any time, a new version of code changes can be submitted by the author which restarts the \cor\ process.

One of the key differences between \ger\ and \phab\ is how changes
are merged when accepted.
For \ger , the default policy is to start
the merging process \emph{automatically} once the acceptance criteria are met.
By default, the reviewer assigning \textquote{Code-Review+2} label to the
\cor\ will enable the automatic merging to take place.
Each project can be also customized further to require a certain
amount of acceptance with certain thresholds from specific people~\cite{mcintosh_2016}.
For \phab, either the author of code changes or someone who has an
appropriate set of permissions will \emph{manually} need to initiate the merging process.

\ger\ and \phab\ use different terminology.
\ger\ uses \textquote{change} to refer to a set of code modifications proposed
in the current iteration (referred to as \textquote{patch set}) of the \cor.
\phab\ uses \textquote{differential revision} which in practice gets shortened to a \textquote{diff}.
The process of applying reviewed changes to the target branch is called \textquote{submitting} in \ger\ and \textquote{landing} in \phab.
For the purposes of our paper, we will refer to that step as merging.

\section{Study Design}

\subsection{Choice of data}
Tools that support sophisticated \cor\ data analytics are widely used in industry. 
For example, Facebook uses \phab, Google utilizes Critique, and Microsoft conducts reviews via CodeFlow.
Since the companies using such tools mainly follow a closed-source model, their data is not available for analysis.

For our study, we sought data from \oss\ projects with the following characteristics:
\begin{enumerate*}[label=(\roman*),before=\unskip{ }, itemjoin={{, }}, itemjoin*={{, and }}]
    \item active multiyear development history
    \item a sizeable number of core contributors
    \item popularity and relevance in their respective fields.
\end{enumerate*}
Suitable projects must also formally track \ttfr, \tta, and \ttm\ using their respective \cor\ environments.
In addition to the above, we selected projects that represent various software abstraction layers and usage scenarios.
This includes a browser, an operating system, a compiler toolset, an office suite, and a \textsc{3D} graphics creation suite.

Historically, the popular communication medium for conducting
\cor s in the \oss\ community has been a mailing list.
Some well-known example projects are Apache, Linux, OpenBSD, and PostgreSQL.
However, mailing lists provide only an approximate means to gather acceptance related data captured semi-formally~\cite{jiang_will_2013,bird_2016}.
For instance, recently Linux kernel started using \textquote{Reviewed-by:} tag to formally signify acceptance of patches, but not necessarily the final sign-off~\cite{linux_cr_guidance_2021}.
OpenBSD uses replies such as \textquote{ok @reviewer\_alias} to denote reviewer acceptance of the changes.

\gh\ added the functionality to capture the basic \cor\ workflow in September 2016~\cite{github_code_review_2016}.
However, the adoption rate of this feature on \gh\ is insufficient for large scale analysis.
We picked \num{100} random \gh\ projects and found that only \num{4} of them
used the \cor\ workflow and even then, the usage pattern was inconsistent.

Our analysis is based on the publicly accessible data mined from \ger\ and \phab.
We selected all available projects in the \ger\ dataset~\cite{yang_2016,yang_2021}.
Here, Eclipse, LibreOffice, and OpenStack are individual projects while GerritHub (analogous to \gh) is a collection of random projects.
The selection, extraction, and cleaning of \ger\ data is described in the original paper~\cite{yang_2016}.
\ger\ projects can have different \cor\ policies (e.g., interpretations and use of labels, and number and role of reviewers).
For example, a project may require an acceptance from two code reviewers instead of one~\cite{openstack_cr_guidance_2021}.
Therefore, when calculating metrics such as \tta,
the policy specific to a project needs to be considered.

For \phab, we mined the full history of \cor s for Blender, \fbsd, \llvm, and \mo.
The selection of \phab\ projects is based on the list of publicly available projects
maintained by \phab\ community~\cite{phabricator_usage}.
We included all the projects with at least \num{10000} \cor s accessible to us.

\subsection{Selection and elimination criteria}
\label{subsec:selection_and_elimination}
To select \cor s that can offer meaningful insights, we applied one selection and many elimination criteria.
We selected \cor s which have undergone complete \cor\
life cycle, i.e., they have been published for review, accepted, and eventually merged into the code base.
Our goal is not to investigate why \cor s do not get accepted or quantify the overall productivity lost by either producing or reviewing the \cor s which never end up being merged.
These criteria imply that we removed code reviews that are committed without acceptance since we cannot measure \tta\ for such reviews.
We also ignored rejected code reviews or code changes without any activity.
Typical reasons for the lack of activity are the \cor\ either being abandoned by the author or ignored by the community.
We removed \cor s without any changes, meaning no source code files or lines of code modified (e.g., binary files).
Finally, to ensure that we are looking at valid data, we selected \cor s where timestamps
follow the logical progression, i.e., \tta\ comes after \cor\ creation time, and \ttm\ after \tta.
We removed the \cor s where for example a reviewer is added, changes are accepted and
cherry-picked at the same moment.\footnote{\url{https://gerrit.libreoffice.org/c/libpagemaker/+/9496}}
This type of pattern indicates that at least formally no substantive \cor\ took place.

Next, we manually inspected the data to look for patterns that otherwise can render the findings meaningless. 
We found a nontrivial amount of \cor s where the author
was also the reviewer, and the changes were \textquote{self-accepted}.\footnote{\url{https://reviews.llvm.org/D20787}}\footnote{\url{https://gerrit.libreoffice.org/c/core/+/7373}}
For \ger\ approximately 18\% of \cor s were self-accepted.
For \phab\ the number was significantly smaller with approximately 0.19\% of \cor s exhibiting this behavior.
Some projects allow such a behavior if the author is a committer and also a long-time contributor.
For the purposes of our study, we do not consider these \cor s to be a
valid representation of a typical \cor\ and exclude them.

Finally, to make sure that a real person is involved, we excluded \cor s where the \emph{only participants}
are the author and automated bots such as various \textsc{CI} related jobs.
To exclude automated bots, we first filtered out the known accounts (e.g.,
\textquote{Jenkins},
\textquote{LibreOffice gerrit bot},
\textquote{RDO CI Service Account}) and then manually inspected the name
of each user account and their activity.
This filter also helps us avoid situations where events such as the start of validation builds,
or the status of sanity checks on the code changes counts as a first response or a substantive comment.

Our initial raw dataset had a total of \initialcrcount\ code reviews which reduced to \finalcrcount\ after applying the above selection and elimination criteria.
For \ger\ projects, we started with \initialgercrcount\ \cor s (conducted from 2013 to 2015).
After applying all the filters, the resulting dataset contains \filteredgerreviews\ \cor s.
For \phab, the initial set contains \initialphabcrcount\ \cor s (conducted from 2012 to 2021)
and final set \filteredphabreviews\ \cor s.

\subsection{Data extraction}
The \ger\ dataset is distributed in the form of a {MySQL} database dump~\cite{yang_2021}.
We developed custom \textsc{SQL} queries (based on the database schema) to analyze the data necessary for our study.
To mine the publicly accessible \cor\ data from the various instances of
\phab, we used Phabry~\cite{cotet_2019,phabry_2021}.
Phabry outputs \cor\ transaction history in a \textsc{JSON} format which we
parse using a custom tool developed in {C\#}.
The resulting output is in plain text format\footnote{\protect\url{https://figshare.com/s/892a48c4fc86bc158e05}} that is analyzed and processed using {R} scripts.

\section{Non-Productive Time in Code Review}
\label{subsection:rq1}

\subsection{Overview of data}
\begin{table*}[ht]
  \caption{Descriptive statistics about \ger\ (first four entries) and \phab\ projects. \textsc{SLOC} is the size of the \cor. Various times (\ttfr, \tta, and \ttm) are given in hours. N = total number, M = mean, Mdn = median, SD = standard deviation.}
  \label{tab:descriptive_stats}
  \begin{tabular}{lrrrrrrrrrrrrr}
    \toprule
    &&\multicolumn{3}{c}{SLOC}&
    \multicolumn{3}{c}{Time-to-first-response}&
    \multicolumn{3}{c}{Time-to-accept}&
    \multicolumn{3}{c}{Time-to-merge}\\
    \cmidrule(lr){3-5}
    \cmidrule(lr){6-8}
    \cmidrule(lr){9-11}
    \cmidrule(lr){12-14}
    Project&\textit{N}&\textit{M}&\textit{Mdn}&\textit{SD}&\textit{M}&\textit{Mdn}&\textit{SD}&\textit{M}&\textit{Mdn}&\textit{SD}&\textit{M}&\textit{Mdn}&\textit{SD}\\
    \midrule
Eclipse & \num{2259} & 818.94 & 32.00 &  13319.63 &  125.95 & 16.17 &  424.27 &  210.39 & 25.52 &  641.07 &  231.82 & 36.58 &  661.86\\
GerritHub & \num{16509} & 601.87 & 27.00 &  13993.87 &  42.84 & 2.87 &  158.95 &  94.24 & 14.16 &  282.53 &  116.64 & 20.86 &  315.20\\
LibreOffice & \num{11567} & 741.60 & 18.00 &  14991.28 &  60.24 & 15.82 &  154.32 &  95.30 & 20.77 &  279.91 &  97.63 & 21.33 &  282.51\\
OpenStack & \num{134391} & 223.65 & 15.00 &  5268.01 &  51.81 & 4.28 &  203.87 &  291.80 & 52.83 &  769.47 &  307.44 & 62.21 &  790.01\\
    \midrule
Blender & \num{6365} & 386.34 & 24.00 &  8133.78 &  126.05 & 8.36 &  654.99 &  326.90 & 19.86 &  1649.03 &  503.46 & 63.21 &  2057.73\\
FreeBSD & \num{14820} & 254.72 & 17.00 &  3568.09 &  101.84 & 4.35 &  634.62 &  226.04 & 9.61 &  1113.61 &  536.00 & 58.31 &  2135.51\\
LLVM & \num{66221} & 283.25 & 51.00 &  5296.55 &  83.99 & 5.53 &  525.88 &  255.32 & 23.48 &  1138.42 &  377.64 & 69.63 &  1433.68\\
Mozilla & \num{97911} & 346.13 & 26.00 &  6255.02 &  43.28 & 5.47 &  182.90 &  83.77 & 14.32 &  319.46 &  190.91 & 46.49 &  641.62\\
    \bottomrule
  \end{tabular}
\end{table*}

We report statistically significant results at a $p < .001$ and use \textsc{APA} conventions~\cite{apa}.
The distribution of the \ttfr\ ($W = 0.13, p < .001$), \tta\ ($W = 0.24, p < .001$),  and \ttm\ ($W = 0.24, p < .001$) of the projects hosted on Gerrit and Phabricator is right skewed (insights from Shapiro-Wilk tests~\cite{shapiro} applied on the data in \cref{tab:descriptive_stats}). 
For our non-normally distributed code review time data, we use non-parametric tests (see~\Cref{subsubsec:metrics}).

When comparing the distinct phases of the code review time between \ger\ and \phab, we notice that:

\begin{itemize}
   \item The median \ttfr\ for \ger\ \cor s ($Mdn$ = 4 hours) is one hour lower than the \phab\ \cor s ($Mdn$ = 5 hours).
   The difference is statistically significant as measured using the Mann-Whitney \textit{U} test~\cite{mann_whitney} $(N_{\ger} = N_{\phab} = \num{100000}; z = \num{-10.34}; p < .001$).
   Here, $N$ indicates the sample size of Gerrit and Phabricator code reviews respectively, $z$ refers to the z-score followed by the p-value indicating statistical significance. 
  \item The median \tta\ for \ger\ \cor s ($Mdn$ = 42 hours) is 26 hours higher than the \phab\ \cor s ($Mdn$ = 16 hours). The result is statistically significant with $z = \num{84.11}$ and  $p < .001$.

    \item The median \ttm\ for \ger\ \cor s ($Mdn$ = 49 hours) is 3 hours lower than \phab\ \cor s ($Mdn$ = 52 hours) with $z = \num{-5.87}$ and $p < .001$.

\end{itemize}

Previous studies have provided insights on code review time of \fbsd\ and \mo\ projects~\cite{zhu_2016,nurolahzade_2009,souza_2015}. 
A study on approximately \num{25000} \cor\ submissions in \fbsd\ via mailing lists showed that the median resolve time (comparable to \ttm) is
\num{23} hours and approximately \num{6} hours for the \ttfr~\cite{zhu_2016}.
In comparison, our data shows a median \ttm\ of \num{54}
hours and median \ttfr\ of \num{4} hours (see \cref{tab:descriptive_stats}).
Another study on \mo\ about how fast \cor s are submitted as a response to bugs of different priority shows that, irrespective of the bug report priority, patches received a review within 24 hours~\cite{nurolahzade_2009}.
One more point of comparison is a study about \num{39770} Firefox
patches created between 2009 and 2013~\cite{souza_2015}.
The study shows that it generally takes \num{137} hours (median) from the time of submission until commit.
Our data covers the time starting 2017 when \mo\ project started using \phab.
It indicates that the recent Firefox \cor s have become shorter.

\subsection{Time prior to first response}

Time-to-response (median) for both \ger\ and \phab\ is less than a day and, in most cases less than 8 hours (see \cref{tab:descriptive_stats}).
To understand the potential reasons for delay in the first response,
we picked \num{100} random \cor s from both \ger\ and \phab .
We applied stratified
sampling considering the number of \cor s in each project where the
\ttfr\ exceeds a typical eight-hour workday.
Inspecting the history of these \cor s reveals that only in 2\% of the 
cases the reviewer expressed a statement related to why their
response was delayed.
In most cases reviewers stated that they were
occupied with something else, but no further details are provided.
It is entirely feasible that the review itself was performed very efficiently,
though the start of the reviewing process was delayed.
Determining what exactly happened before someone decided to eventually respond to a
\cor\ is not possible because such data is not tracked anywhere.
Since we cannot determine reasons for delay in \ttfr,
we limit ourselves to suggest that it is desirable to shorten the \ttfr\ and
reduce the number of iterations a \cor\ goes through.

Projects Eclipse and LibreOffice are two notable exceptions for the median \ttfr\  by taking approximately twice the time. 
We manually inspected the history of \num{100} randomly selected \cor s from each project using stratified sampling.
We did not find a reason that can explain why median \ttfr\ for these projects is higher.
That said, there can be several confounding factors (e.g., number of available reviewers~\cite{yu_determinants_2016,yu_wait_2015},
affiliation of authors~\cite{baysal_2016, sadowski_modern_2018}, and technical infrastructure~\cite{hilton_2016,yu_determinants_2016,yu_wait_2015}) that can potentially explain the difference. 

\subsection{Post-acceptance time}
Referring to the time when a code change could have been merged and when the actual merge happens (see the time between \textquote{Accepted} and \textquote{Merged} in \cref{fig:accepted_cr_timeline}), 
we notice that \phab\ projects spend a large
part of the entire \cor\ lifetime in the post-acceptance state in comparison to Gerrit (see \cref{tab:wasted_percentage_post_accept} ). 
The percentage of the total \cor\ lifetime accounting for the time between acceptance and merge for \ger\ \cor s ($Mdn$ = 1.00\%) is lower than for \phab\ \cor s ($Mdn$ = 43.00\%).
A Mann-Whitney \textit{U} test indicated that this difference is statistically significant $U(N_{\ger} = \num{100000}, N_{\phab} = \num{100000}) = \num{6735634167.5}, z = \num{-255.08}, p < .001$.
This finding represents a noticeable opportunity to decrease \ttm\ in \phab\ \cor s.

\begin{boxed_new}
Depending on a \phab\ project, between 29\% and 63\% of the overall \cor\ lifetime is spent waiting for the accepted changes to be merged (see \cref{tab:wasted_percentage_post_accept}).
\end{boxed_new}

\begin{table}[ht]
  \caption{Percentage of total \cor\ lifetime accounting for the time between acceptance and merge. N = code review count, M = mean, Mdn = median, SD = standard deviation.}
  \label{tab:wasted_percentage_post_accept}
  \begin{tabular}{llrrrr}
    \toprule
    Project&Review tool&\textit{N}&\textit{M}&\textit{Mdn}&\textit{SD}\\
    \midrule
Eclipse & Gerrit& \num{2259} & 11.67\% & 0.09\% &  26.00\%\\
GerritHub & Gerrit& \num{16509} & 18.42\% & 0.63\% &  31.82\%\\
LibreOffice & Gerrit& \num{11567} & 2.30\% & 0.01\% &  12.55\%\\
OpenStack & Gerrit& \num{134391} & 11.06\% & 1.00\% &  22.50\%\\
\midrule
Blender & Phabricator&\num{6365} & 40.28\% & 29.00\% &  37.59\%\\
FreeBSD & Phabricator&\num{14820} & 55.58\% & 63.00\% &  37.71\%\\
LLVM & Phabricator&\num{66221} & 41.66\% & 32.00\% &  36.52\%\\
Mozilla & Phabricator&\num{97911} & 48.81\% & 48.00\% &  36.21\%\\
    \bottomrule
  \end{tabular}
\end{table}

Switching to Gerrit, most \ger\ \cor s are merged in a matter of minutes after acceptance.
The only exception is the {O}pen{S}tack project which at some point in the
project history started using synchronous validation builds that must complete before the changes are merged.
Therefore, the time it takes to merge the {O}pen{S}tack code changes
is directly dependent on the duration of the build.
Other than the exceptional behavior of the {O}pen{S}tack project, the lower percentage difference between \tta\ and \ttm\ for Gerrit projects was expected. 
We believe that this small difference is attributed to the automatic merge policy of the accepted changes in \ger .

We further investigated the root cause for the delays in \ger\ projects, where the merge time exceeded \num{5} minutes (more than the median time). 
As before, we randomly picked \num{100} \cor s using stratified sampling and explored the \cor\ history for the causes.
We observed that the main causes of delay relate to merge conflicts,\footnote{\url{https://git.eclipse.org/r/c/platform/eclipse.platform.ui/+/2311}}\footnote{\url{https://gerrit.libreoffice.org/c/core/+/9133}} build failures\footnote{\url{https://git.eclipse.org/r/c/cdt/org.eclipse.cdt/+/2310}}\footnote{\url{https://review.opendev.org/c/openstack/python-swiftclient/+/91524}} or the time it takes
to complete the verification build(s).\footnote{\url{https://gerrit.libreoffice.org/c/core/+/27133}}\footnote{\url{https://review.opendev.org/c/openstack/oslo.i18n/+/92678}}
Only in 7\% of the cases the reason for the delay was not explicitly clear.
Based on our industry experience with \textsc{CI} systems,
we suspect
that typical infrastructure related delays such as intermittent networking issues or
increased wait times for scheduled jobs can be other possible explanations. 
Although, we do not have logs from the \textsc{CI} systems to verify these theories. 

In this section, we identified \npt\ in \cor s. Next, we explore the characteristics of the \cor\ that can contribute to the \npt.

\section{Contributors of Non-Productive Time}
\label{subsection:rq2}

\subsection{Time-to-first-response}
An intuitive approach to reduce the overall \ttm\ is to minimize the number of
\cor\ iterations.
In an ideal case, the initial published set of changes will be immediately
accepted in as little time as possible.
Acceptance of changes being associated with a first response is a desired outcome
to increase the code velocity.
The time it takes for a reviewer to respond to someone's
proposed changes is dependent on several factors,
only a few of which an author can directly control.

\begin{figure}[!t]
    \centering
    \includegraphics[width=0.4\textwidth,keepaspectratio]{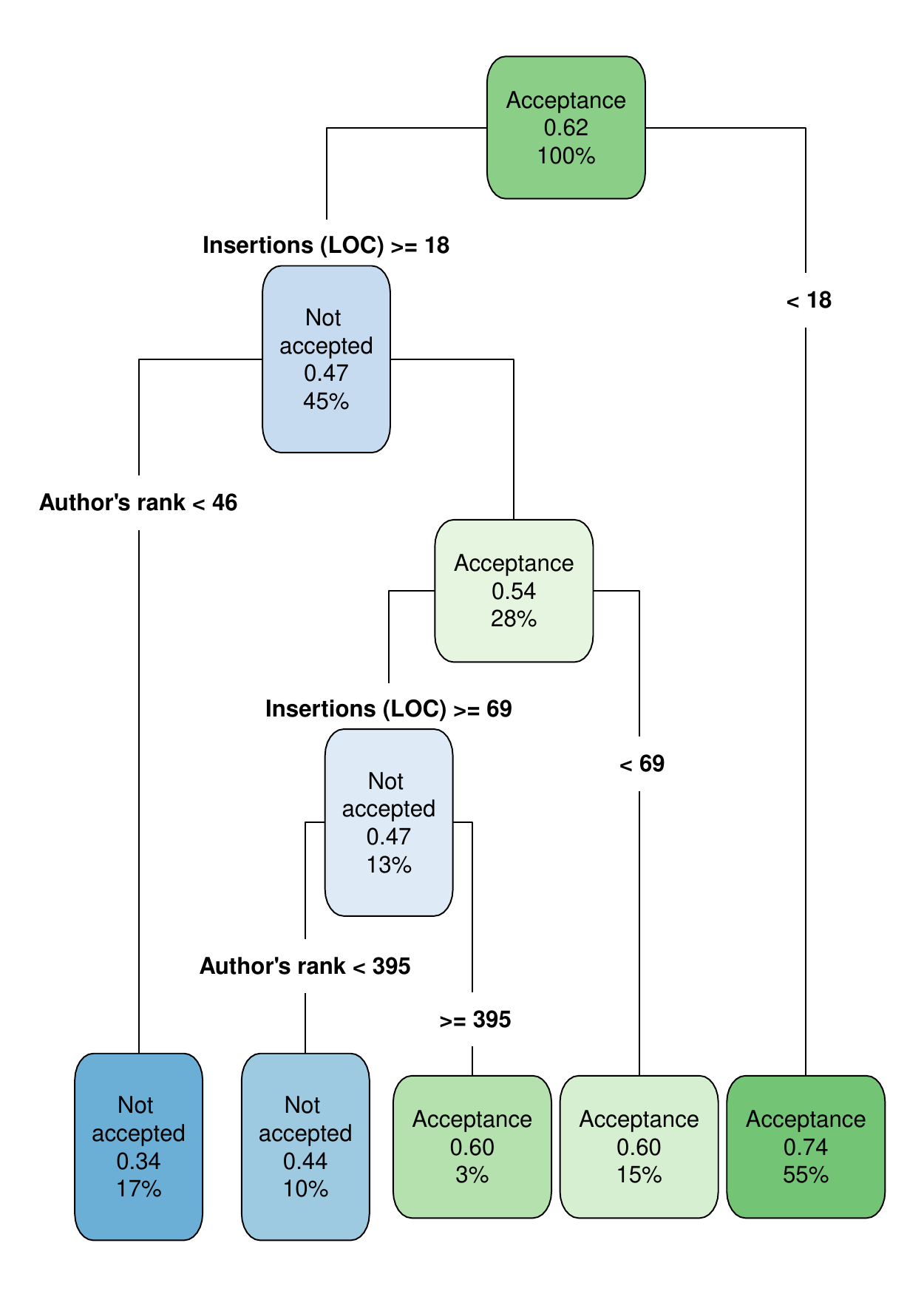}
    \caption{Decision tree for determining if the first response to a \phab\ \cor\ is an acceptance. Author's rank is a total number of \cor s the author of changes has successfully merged for a particular project at the time of the submission of a current \cor.}
    \label{fig:decision_tree}
\end{figure}

Studies show that a code review is characterized by size and composition~\cite{baysal_2016,izquierdo-cortazar_2017,ram_what_2018}. 
It is also linked to the identity and reputation of the author and the reviewer(s)~\cite{kononenko_2018,pinto_who_2018}.
To understand the factors influencing the first response time, we build a decision tree using the above factors as input. 
\cref{fig:decision_tree} shows how the key variables influence the first response time. 
The accuracy of the resulting decision tree is 68\%.

The analysis above is based on \phab\ data for which we have a complete \cor\ history
containing all the changes in the chronological order.
The dataset from \ger~\cite{yang_2016} contains only a subset of the \cor s for each project.
That makes the \ger\ data unsuitable for calculating the author and reviewer ranks
because to perform valid calculations we need a consecutive timeline for all the \cor s.
Another issue with the \ger\ dataset is that it does not formally distinguish between various \cor\ roles such as
Author, Committer, Owner, Uploader~\cite{gerrit_roles}.
As a result, we are not able to determine who \emph{exactly} authored the code
changes versus just committed them or had permissions to upload changes
to the \cor\ tool.

\begin{boxed_new}
    Decision tree indicates that smaller code changes
    ($< \num{18}$ SLOC of new code) have a higher chance of immediately
    being accepted.
    For changes larger than that author's rank ($\geq 46$ previously
    completed \cor s) becomes a factor.
\end{boxed_new}

\begin{figure}[!t]
    \centering
    \includegraphics[width=0.4\textwidth,keepaspectratio]{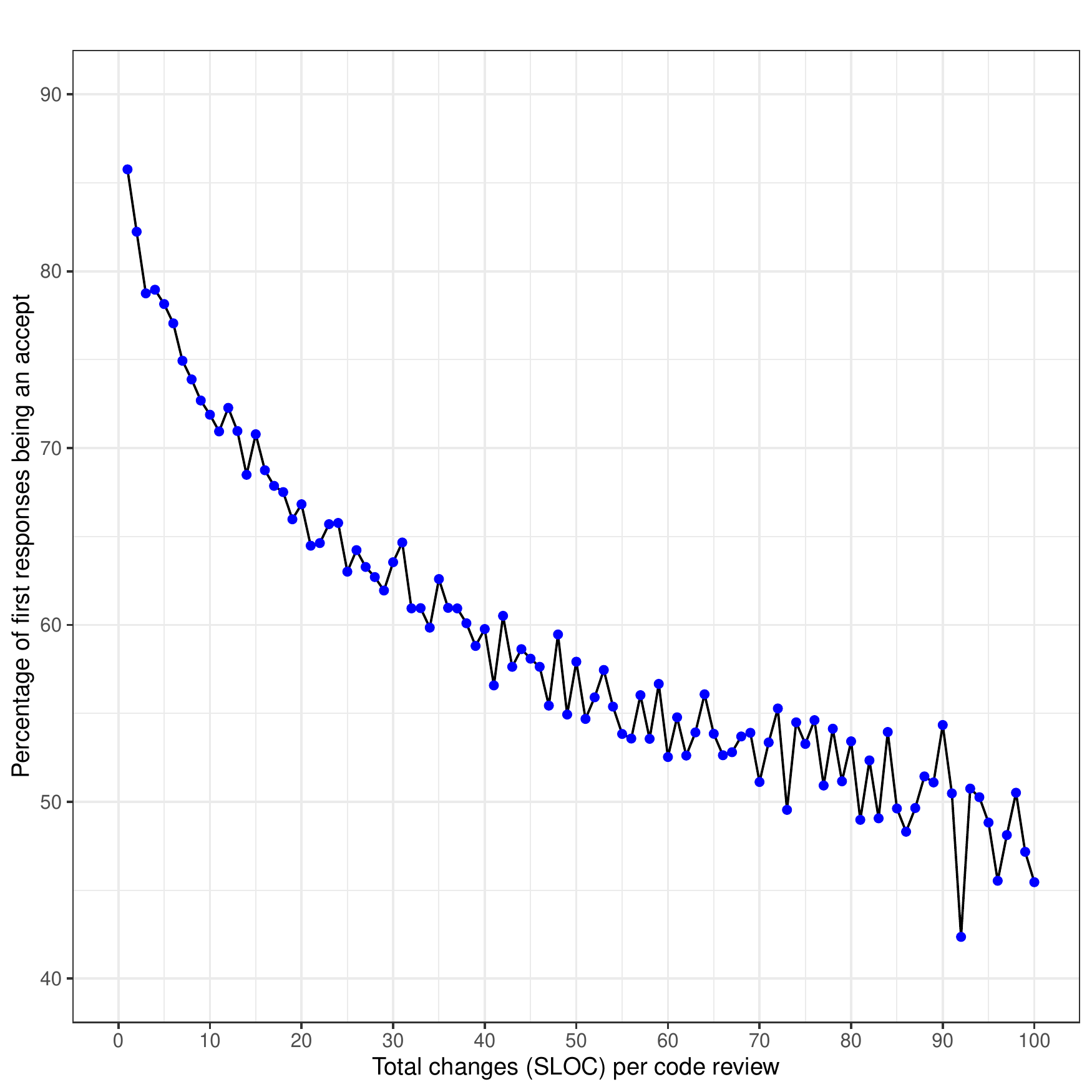}
    \caption{Percentage of first responses being an acceptance for \phab\ \cor s containing [1..100] \textsc{SLOC}.}
    \label{fig:per_line_ttfr_phabricator}
\end{figure}

\begin{figure}[!t]
    \centering
    \includegraphics[width=0.4\textwidth,keepaspectratio]{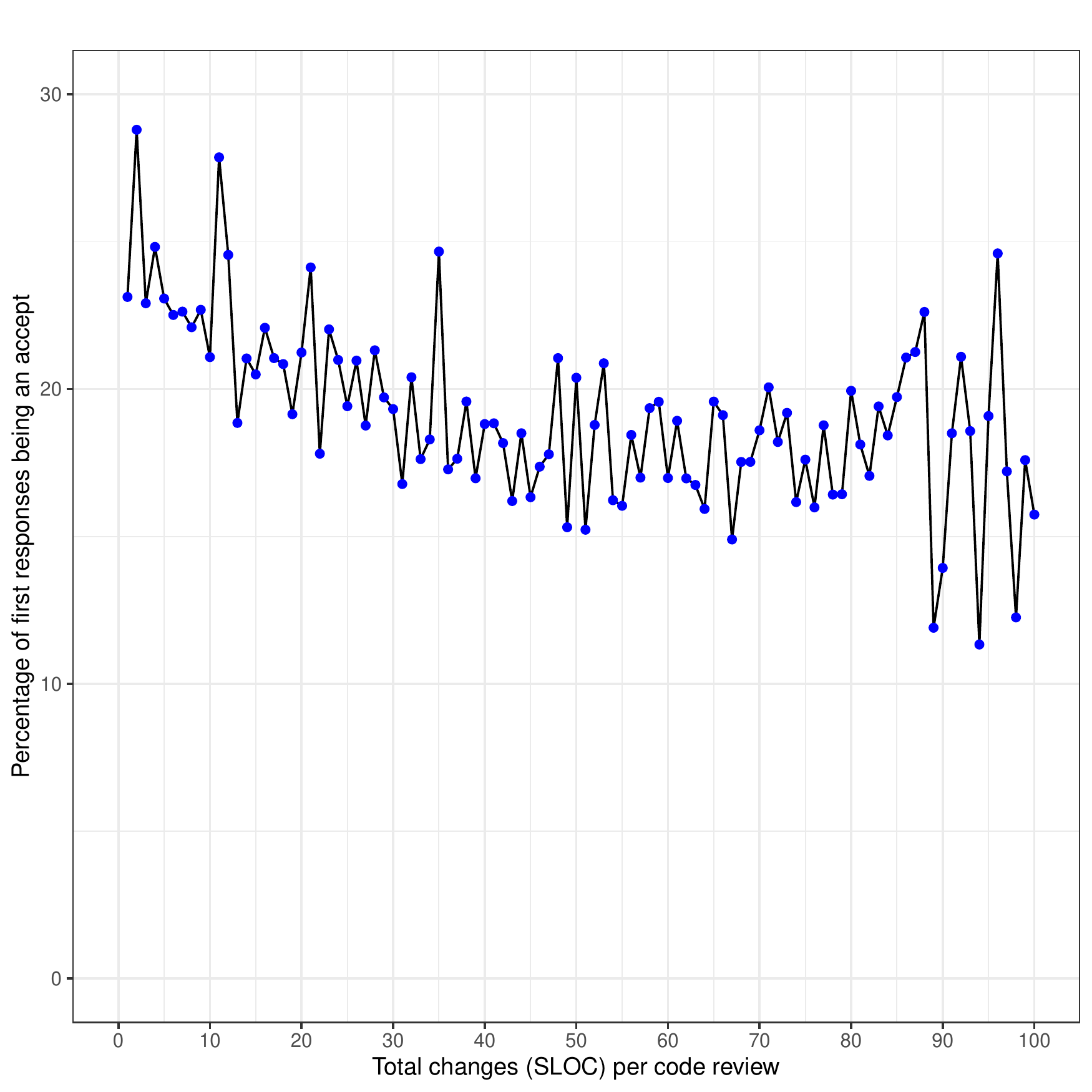}
    \caption{Percentage of first responses being an acceptance for \ger\ \cor s containing [1..100] \textsc{SLOC}.}
    \label{fig:per_line_ttfr_gerrit}
\end{figure}

The results from the decision tree direct us towards investigating if there
is a trend associated with the size of the \cor s in \textsc{SLOC} and the
percentage of \cor\ request being accepted as part of the first response.
We focus on the \cor s with a size of [1..100] \textsc{SLOC}.
The subset of \cor s with a size up to \num{100} \textsc{SLOC} accounts for approximately
75\% of all the \phab\ and 79\% of all the \ger\ \cor s.
We can see in \cref{fig:per_line_ttfr_phabricator} that for \phab, the
percentage of acceptance decreases as the size of \cor\ increases.
Per \cref{fig:per_line_ttfr_gerrit} we observe that there is similarly a decreasing trend for \ger,
but it is less pronounced and has more variation.
The percentage of first responses being an 
acceptance stays in a fixed range regardless of the \cor\ size.
Given that in \ger\ acceptance translates into an immediate merge request,
we speculate that reviewers are more mindful of the impact
their decision has.
Therefore, only the \cor s in which the reviewer has an utmost confidence
in are accepted during the first iteration.
In contrast, with \phab, both the author and reviewers have a \textquote{safety buffer}
during which time they can rethink the merging decision.

\begin{figure}[!t]
    \centering
    \includegraphics[width=0.4\textwidth,keepaspectratio]{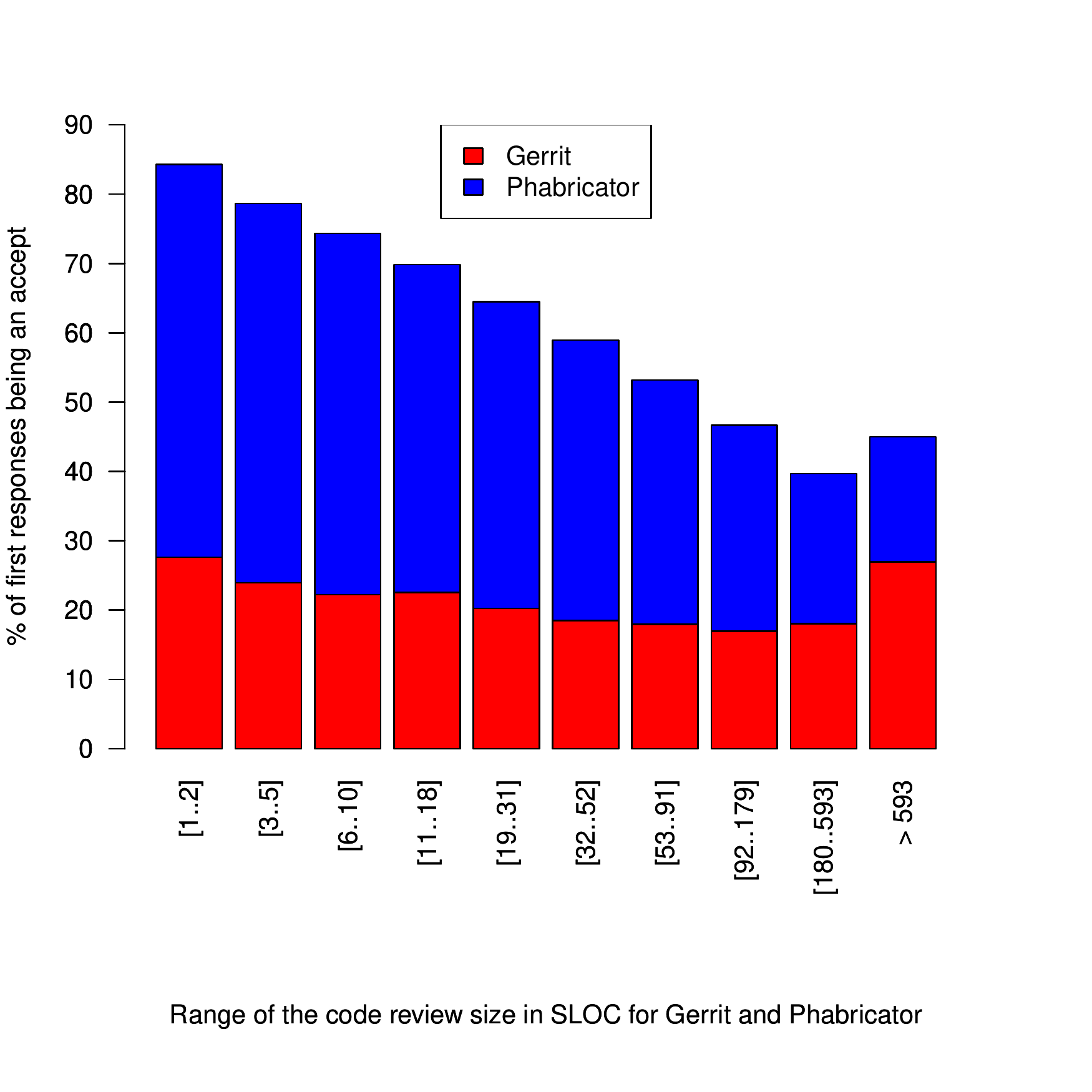}
    \caption{Percentage of first responses being an acceptance for \ger\ and \phab\ \cor s for different size buckets. Each bucket contains 10\% of the population.}
    \label{fig:barplot_ttfr_both}
\end{figure}

To investigate the overall trend, we divided the existing \cor s into buckets of different ranges and calculated the percentage of first response being an acceptance.
Each bucket contains approximately 10\% of the overall population.
The results are presented in \cref{fig:barplot_ttfr_both}.
An interesting observation from the acceptance pattern is that $\geq 180$ \textsc{SLOC}, the percentage of
immediate acceptance increases forming a bathtub curve~\cite{bathub_curve_1, bathtub_curve_2}.
Based on our industry experience, one possible explanation for this behavior is that beyond a certain threshold, engineers tend to review bigger code changes
lightly and accept them at a rate comparable to smaller changes.
There can be multiple explanations for this observation. 
One of which is a harmful practice suggesting that beyond a certain threshold the thoroughness of \cor s decreases.
Contingent on what takes place here, which the future studies should verify, using tools to automatically decompose proposed changes into smaller
\cor s can be a potential mitigation technique~\cite{barnett_helping_2015}.

\subsection{Post-accept interval}
\label{subsubsec:post-accept-interval}
This section is based on \phab\ \cor s only because the default \ger\ workflow
immediately triggers a merge attempt after changes have been accepted.
Our stratified sample of random \num{250} \ger\ \cor s indicates that merge conflicts or build failures
happen in less than 1\% of cases.
We use the classifications below to describe what happens to differential
revisions in \phab\ after they have been accepted.
To understand the intent behind these actions we manually analyze a
sample of random \num{100} instances from each category.
Based on the results from the analysis we classify each type of
post-accept behavior as a negative, positive, or neutral indicator.

\begin{itemize}
    \item \textbf{No action}: There was no \cor\ related activity 
    triggered by a human between
    the acceptance and until the changes were eventually merged.
    We classify this as a positive event since this time can be reduced for increasing code velocity.
    \item \textbf{Changes requested}: The original or a new
    reviewer requested additional changes
    to the code, thereby invalidating the original acceptance.
    For decreasing \cor\ lifetime we consider this to be a negative event.
    However, when it comes to code quality, this can be a positive
    development because a reviewer either found defect(s) which were not
    discovered before, realized that earlier acceptance was submitted
    by mistake or found something else which is significant enough
    to delay merging the code changes.
    \item \textbf{Diff updated}: refers to updated code changes. This action
    could indicate anything from author performing extra validation,
    resolution of merge conflicts or applying some optional feedback from
    the original \cor. We consider this a neutral event.
    In majority of the cases the two main reasons behind this event are
    either the author making supposed improvements to the code or resolving
    merge conflicts.
    \item \textbf{Inline comments}: mean that someone commented on specific parts
    of code. Typically, this means either optional suggestions, something
    which was missed during the original \cor\ (though not serious enough
    to warrant requesting changes) or, in some cases
    even positive feedback about code changes.
    We consider this to be a semi-negative event. %
    \item \textbf{Comments}: refer to arbitrary text comment about the \cor. This
    can have a variety of meaning ranging from a critique to thanking the
    author to an emoji to a reference from or to another \cor.
    We consider this to be a neutral event.
\end{itemize}

In practice, we observe that each \cor\ can have multiple events from each category associated with its
post-acceptance.
For example, an author can update accepted changes,
another engineer can make a comment on the code,
automated bot may attempt to apply changes to the latest version of source tree, etc.
A percentage distribution of the post-accept behavior is presented in \cref{tab:post_accept_behavior}.

\begin{boxed_new}
Once the changes are accepted, only in a small number of cases acceptance is subsequently invalidated (additional changes requested).
For half of the \cor s there is no associated post-accept activity
and changes are merged as is.
\end{boxed_new}

\begin{table}[ht]
  \caption{Percentage of different post-accept behaviors per each \phab\ project.}
  \label{tab:post_accept_behavior}
  \begin{tabular}{lrrrrr}
    \toprule
    Project&        No&  Change &   Diff &      Code& Com-\\
            &activity & request & update & comments & ments\\
    \midrule
Blender & 59\% & 4\% & 19\% & 17\% & 31\%\\
FreeBSD & 56\% & 1\% & 16\% & 22\% & 35\%\\
LLVM & 45\% & 1\% & 22\% & 31\% & 44\%\\
Mozilla & 42\% & 2\% & 46\% & 27\% & 41\%\\
    \bottomrule
  \end{tabular}
\end{table}

\section{Discussion}

\subsection{Interpretation of our results}

One of the first questions that should be answered is whether the period
we consider non-productive is wasteful.
\mcr\ is an asynchronous process by its nature.
Engineers are not per se blocked from working on any other tasks
till an event requiring
their attention (e.g., \cor\ feedback or acceptance, merge conflict
with someone else's code changes, failing build, etc.) occurs.
However, the \cor\ request itself is in an idle state and ideally
the duration of that state should be reduced.
If we want the code changes to be reviewed and merged
in \num{24} hours or less,
eliminating idle waiting enables us to significantly reduce the \ttm.
The trade-off between merging potentially incorrect code versus increasing
code velocity may depend on what a particular project values the most.
The data in \cref{tab:post_accept_behavior} shows that
after acceptance, new changes are requested on average only in 2\% cases.
Half of the \cor s have no post-accept activity associated with them.

Reducing the difference between \tta\ and \ttm\ may be
as easy as modifying a project's \emph{default} configuration settings
to merge code changes automatically.
Modifying the default \cor\ policy
(and if needed then promptly reverting those changes) 
for projects enables quick experimentation.
Tools like \ger\ and \gh\ enable configuring a custom set of rules
per repository or project~\cite{github_auto_merge_2021,gerrit_2021}.
Those rules specify a set of conditions under which a commit is automatically
merged.
A default \ger\ rule is to attempt to merge changes when a \cor\ has
received at least one \textquote{Code-Review+2} response.
For \gh\ the \textquote{allow auto-merge} option can be enabled.
Engineers can then configure a custom set of branch protection rules to determine
the required number of reviewers and checks that code changes need to pass
before the automatic merge is attempted.
The public version of \phab\ does not support automatically landing
the accepted changes as of writing this paper.

Changing the merge policy may have unintended negative consequences.
For the reviewer, automatic merging means that acceptance implies that
code changes will be merged as is.
Even the minor or optional feedback will require another iteration
of updating the changes and accepting them again.
Extra iteration(s) in turn may lengthen the \ttm\ which is the
opposite of our desired outcome.
For the author of changes the \textquote{safety buffer}, during which
the author can reconsider merging the changes, make last minute updates
or ask for feedback from additional reviewers, disappears.
Additional risk increases the demands which are placed on verifying
that the initial set of changes are correct by executing (potentially unnecessary)
verification steps or delaying publishing the \cor\ request.

Optimizing \ttfr\ is a far more complex problem with many confounding
variables and will require a separate study to be conducted.
From our findings we can see those two variables (author's rank and the
number of lines of new code inserted) are key decision points in the
decision tree.
This feels right intuitively.

\subsection{Implications for research and practice}

\begin{enumerate}
    \item \emph{Commit-on-accept model increases code velocity}.
One of the core implications of our study is that the choice of the \cor\
infrastructure and corresponding \cor\ policy matters.
In systems like \ger, the merging and committing of changes is done
automatically.
Similar policy can be configured for \gh\ projects as well.
We think that given the findings presented in \cref{subsection:rq2}
and given the low rate of errors on initial acceptance,
the commit-on-accept model may be beneficial.
    \item \emph{Focus on \ttm\ versus \tta}.
Most of the existing papers focus on \tta\ which from practitioner's point
of view is just an intermediate step until the point when code changes
become verified and usable.
Historically that focus may have been justified because there was no significant
difference between these points in time.
With the addition of modern \textsc{CI} pipelines and increased usage of
single repository and \textquote{trunk based development} in industry,
the process of getting the changes merged is more involved and time-consuming than a
simple commit command~\cite[p.~339]{winters_2020}.
    \item \emph{Adopting bots to notify engineers of pending \cor s}.
One of the simplest explanations for why engineers do not react immediately
to \cor s is because they are not aware that an action is expected from them.
Most of the existing \cor\ systems enable triggering notification
emails.
Based on our industry experience, direct requests coming from an engineer with
whom the reviewer does not have an existing working relationship may
be considered both as \textquote{nagging} and rude, and result in an
opposite desired effect.
A study investigating two projects which used \cor\ bots for a couple
of years found that the number of pull requests merged (monthly) increased after the adoption of a \cor\ bot~\cite{wessel_2020}.
Automated bots can act as a neutral mechanism reminding engineers
of pending tasks.
    \item \emph{Detailed understanding of the composition of \textsc{CI} pipeline}.
To determine what steps in the \cor\ process can be optimized,
we first need to understand how
different parts contribute to the overall \ttm.
The contents of the \textsc{CI} pipeline may highly depend on the project
and organization.
It is advisable that periodically projects use tools applicable to their
platform (e.g., CodeFlow Analytics, Gerrit DevOps Analytics, and
\gh\ Insights) to monitor the performance of \cor s and the time taken during the different
stages of \textsc{CI} process.
\end{enumerate}

\section{Threats to Validity}

Like any other study, the results we present in this paper are
subject to certain categories of threats.
In this section we enumerate the threats to construct, internal,
and external validity~\cite{shull_guide_2008}.

The threats associated with \emph{construct validity} are caused by
not interpreting or correctly measuring the theoretical constructs
discussed in the study.
One of the core assumptions we make is that \cor\ acceptance really
means acceptance, i.e., changes can be submitted as is or with a
trivial number of modifications needed.
We also presume that the reviewer providing an acceptance for a
\cor\ has the authority to grant the permission for the changes to
be made.
\ger\ enables implementing custom \cor\ policies for each project.
We inspected these policies for the projects we studied and took the differences
into account when implementing our data analysis algorithms.
We found only one exception from the default \ger\ \cor\ process.
The {O}pen{S}tack project has a policy requiring two positive reviews from
core reviewers and uses \textquote{Workflow+1} label versus
\textquote{Code-Review+2} label to signify the final acceptance~\cite{openstack_cr_guidance_2021}.

One of the concerns for \emph{internal validity} is the interpretation of the
results and whether the conclusions we present can be drawn from the
data available.
We trust the data coming from the \cor\ tool.
To mitigate the errors present in source data we removed
entries with inconsistent timestamps and also samples
where a \cor\ was clearly not conducted.
For example, self-accepted \cor s and those with zero \tta\ or \ttm.
In addition, we manually sampled hundreds of \cor s and verified
their content and timestamps in \cor\ tool and version control system.

Threats to \emph{external validity} are related to the application of
our findings in other contexts.
Our focus is restricted to the \oss\ projects utilizing \cor\ tools
which enable formally tracking the timestamps capturing both \tta\ and
\ttm.
Though several organizations developing commercial software
(e.g., Facebook, Google, and Microsoft) have embraced
the \oss\ software in recent times, most of their code is
still developed using a closed-source model.
Due to this, as with all empirical studies, the results are not to be generalized outside of the existing context without further replication in other environments and contexts. %

\section{Conclusions and Future Work}

This study quantifies \npt\ in \ger\ and \phab\ \cor s and investigates what happens to the code reviews after they are accepted.
Our study shows that in more than half of the cases no activity is done between acceptance and merge. 
Our exploration into the cause of delay offers actionable insights into potentially increasing code velocity.
We estimate that in case of \phab\ projects the code velocity can be increased by 29--63\%.

From here on, we foresee the following research directions:

\begin{enumerate}
    \item \emph{Reasons why additional code changes are requested after the
initial acceptance}.
Are the instances when the first acceptance is considered
incorrect caused by the initial reviewer missing defects
they noticed later or by another reviewer discovering
additional set of problems with the code?
Is there a relationship between factors such as the size of
changes, previous contribution history of the author,
experience of the reviewer, etc.?
Given that invalidating the initial acceptance is
a failure in the \cor\ process it is important to
understand the etiology of these cases and explore potential
for process improvements.
    \item \emph{Factors influencing the \ttfr}.
Because \ttfr\ is responsible for a significant portion of
overall \ttm\ then shortening this time is the next obvious step
when attempting to increase the code velocity.
Several variables can
impact how fast a qualified reviewer reacts to a \cor .
These variables can include everything from author's identity,
availability of reviewers, complexity and size of changes,
interpersonal relationship between the reviewer and author, etc.
The goal is to understand how a code change author
can solicit a faster response.
    \item \emph{Implications of \cor\ policy on defect density}.
We do not have data points on different \cor\ policies (acceptance being gated by \textsc{CI}
validation results, number of required reviewers per each change, 
committing changes automatically on accept versus manually
by the author, etc.).
One future research direction can be to investigate if differences in policy have a measurable
impact on code quality or defect density.
For example, do projects requiring two reviewers (instead of
one) or mandating that all code changes must be reviewed by
a member of core committers group, have a lower
defect density?
Is the Linus's law stating that \textquote{given enough eyeballs, all bugs are shallow} supported by real evidence~\cite{raymond_cathedral_1999}?
\end{enumerate}

\bibliographystyle{ACM-Reference-Format}
\bibliography{msr-2022-npt}

\end{document}